\documentclass[aps,prb,superscriptaddress,tightenlines,floatfix,twocolumn,notitlepage]{revtex4-1}
\usepackage{graphicx}
\usepackage[english]{babel}
\usepackage{amsmath}
\usepackage{amssymb}
\usepackage{tensor}

\newcommand{\vk}{\mathbf{k}}

\newcommand{\be}{\begin{eqnarray}}
\newcommand{\ee}{\end{eqnarray}}
\newcommand{\p}{\partial}

\newcommand{\dc}{c^{\dagger}}

\def\ket#1{|#1\rangle}
\def\bra#1{\langle #1 |}

\begin{document}

\title{Three dimensional two-band Floquet topological insulator with $Z_2$ index}

\author{Yan He}
\affiliation{College of Physical Science and Technology,
Sichuan University, Chengdu, Sichuan 610064, China}
\email{heyan$_$ctp@scu.edu.cn}

\author{Chih-Chun Chien}
\affiliation{School of Natural Sciences, University of California, Merced, CA 95343, USA.}
\email{cchien5@ucmerced.edu}

\begin{abstract}
We present a class of three dimensional (3D) two-band Floquet topological insulators constructed from two-dimensional Floquet topological insulators with a $Z$ topological index. It is shown that the 3D two-band Floquet topological insulator has a $Z_2$ topological index, whose value can be obtained by numerical calculations or by using a relation to the winding number. The classification of the 3D $Z_2$ Floquet topological insulator, however, cannot be attributed to the stable homotopy groups. Thus, it is an example outside the proposed K-theory classifications of Floquet topological insulators. We also analyze the edge modes of the 3D $Z_2$ Floquet topological insulator and find the parity of the number of edge modes reflects the bulk $Z_2$ index.
\end{abstract}

\maketitle

\section{Introduction}
In recent years, topological properties of materials have been a focus in both theoretical and experimental studies of condensed matter physics~\cite{Kane_TIRev,Zhang_TIRev,ShenTI,Chiu2016}. In addition to topological matter in equilibrium, periodically driven systems can also exhibit non-trivial topological properties \cite{Lindner,Cayssol13,Eckardt17,Oka18}. Interestingly, a static, topologically trivial system may acquire non-trivial topological signature if driven periodically in time. This type of periodically driven topological systems may be realized by shining light on materials~\cite{Cayssol13,Wang13}, varying the parameters periodically in cold-atom systems~\cite{Eckardt17}, photonic systems~\cite{Maczewsky17}, or classical acoustic systems~\cite{Fleury16}.

The time evolution of those periodically driven systems leads one to expect that Bloch's theorem well-known in solid-state physics~\cite{Ashcroft_book} can also be applied to the time direction, just as it can be applied to the spatial directions. Therefore, one can introduce a time-component counterpart of the lattice momentum, which is called the quasi-energy. The result actually was obtained by mathematician G. Floquet \cite{ODE_book} long before Bloch's theorem. When the periodically driven topological systems exhibit band structures (in quasi-energy) similar to the static topological insulators, they are called the Floquet topological insulator (FTI)~\cite{Lindner,Cayssol13,Eckardt17,Oka18}. Properties of stacked 2D FTI have been studied in Ref.~\cite{Ladovrechis18}. In addition to clean systems, there have been recent theoretical~\cite{Titum,Nathan,Fulga16,Tauber18} and experimental~\cite{Hainaut18} studies on FTIs in disordered systems.

Similar to the static Chern insulator, the topological signatures of FTI manifest in two different ways: The first is the topological index from the system with periodic boundary condition. The second is the topological edge modes localized at the boundary when open boundary condition is imposed. For the Chern insulator, there is a bulk-boundary correspondence relating the sum of the Chern numbers of the occupied bands and the number of topological edge modes at the open boundary~\cite{Kane_TIRev,Zhang_TIRev,ShenTI,Prodan_book}. The situations are more complicated for the FTI because the edge modes can exist even when the Chern numbers of all the bands are zero. A bulk-boundary correspondence of FTI has been established in Ref.~\cite{Rudner}. More discussions can be found in Refs.~\cite{Sadel17,Graf18}, including FTIs with disorder. Roughly speaking, the number of edge modes entering a band minus the number of edge modes leaving the same band still equals the Chern number of the band, similar to the counting for the static Chern insulator. The main difference between static Chern insulator and FTI is that the quasi-energy of the latter is by definition a phase angle and thus is periodic in $2\pi/T$, where $T$ is the driving period. Because of this, it is possible for an edge mode to connect the top band and the bottom band~\cite{Bandnote}.

The topological index of FTI can also be obtained as a winding number~\cite{Rudner}. For a 2D FTI with a trivial one-period time-evolution operator $\mathcal{U}(T)=1$ and periodic boundary condition, both the momentum- and time- dependences are periodic. The time-evolution operator then defines a mapping from a 3D torus to the unitary group $U(N)$. Here $N$ is the number of bands. If one ignores the non-trivial cycle of the torus, i.e., treating~\cite{Kennedy15} $T^3$ as $S^3$ in homotopy groups, the topological classification is given by~\cite{Rudner} the homotopy group $\pi_3(U(N))=Z$. This result suggests another class of FTI with the topology of the non-trivial homotopy group~\cite{Puttmann03} $\pi_4(U(2))=Z_2$, whose construction is made possible by extending the 2D FTI to a 3D FTI. We will show the resulting 3D two-band FTI has a $Z_2$ classification similar to the 3D time reversal topological insulator (TI).

Due to the fact~\cite{Puttmann03} $\pi_4(U(N))=0$ for $N\ge3$, the $Z_2$ index can only apply to two-band ($N=2$) models in 3D. Adding more bands will lead to topologically trivial models, so the 3D two-band FTI discussed here is more suitable for low-energy effective models when the focus is on two bands only. Another important subtlety is that, while the static 3D TI can have a $Z_2$ index protected by time reversal symmetry, the $Z_2$ index of the 3D FTI does not require any symmetry. Moreover, the $Z_2$ index of the 3D FTI is different from the index for the 2D time-reversal invariant FTI discussed in Ref.~\cite{Carpentier15}. Similar to the static topological insulator, FTIs of multi-band systems can be classified by the stable homotopy group or K-theory according to their discrete symmetries, such as time reversal or chiral symmetry~\cite{Nathan15,Fruchart16,Roy}. In contrast, the 3D FTI proposed here is due to a low-dimensional homotopy group and may be considered as an exception to the periodic table of FTI. More discussions on the classification of topological indices using homotopy can be found in Ref.~\cite{Kennedy15}.

The framework for constructing the 3D two-band FTI applies to simple models showing piecewise-constant time dependence as well as general models showing explicit time dependence. When open boundary condition is imposed, topological edge modes of the 3D two-band FTI will emerge at the boundary. We will show some examples and discuss the relation between the number of edge modes and the bulk topological index. Moreover, the edge modes of the 3D two-band FTI will be shown to be robust against weak disorder in the onsite potential or hopping coefficient. While the $Z$ index of a 2D FTI is associated with the Chern number of its Hamiltonian~\cite{Rudner}, we found a similar connection between the $Z_2$ index and its Hamiltonian mapping. Therefore, the 3D two-band FTI offers an additional playground for studying topological properties of time-dependent systems.

The rest of the paper is organized as follows. In Section \ref{sec-2d}, we briefly review the 2D FTI and the computation of the winding number. In Section \ref{sec-3d}, we generalize the 2D FTI to the 3D FTI and compute the $Z_2$ index. Then, in Section \ref{sec-edge} we consider some models of the 3D two-band FTI with open boundary condition and discuss the number of edge modes.  A connection between the $Z_2$ index and the parity of the total number of edge modes is also discussed. Section~\ref{sec:conclusion} concludes our work.

\section{2D FTI with $Z$ index}\label{sec-2d}
We briefly review the topological classification of two-dimensional FTI by closely following Ref.~\cite{Rudner}. If a time-dependent, $N$-band Hamiltonian satisfies $H(t+T)=H(t)$, the Floquet theorem requires its eigenstate $\ket{\psi(t)}$ to satisfy
\be \label{eq:eigen}
\ket{\psi(t+T)}=\mathcal{U}(T)\ket{\psi(t)}=e^{i\epsilon T}\ket{\psi(t)}.
\ee
Here we introduce the time-evolution operator 
\be
\mathcal{U}(\vk,t)=\mathcal{T}\exp\Big(-i\int_0^{t} H(\vk,t')dt'\Big),
\ee
and call the time-evolution operator over one period, $\mathcal{U}(T)$, the Floquet operator. Here $\mathcal{T}$ denotes the time ordering. In Eq.~\eqref{eq:eigen}, $\epsilon$ is the quasi-energy, which is the phase angle of the eigenvalues of the Floquet operator. Although the Floquet operator determines the quasi-energy spectrum, to characterize the topology of a 2D FTI, one has to consider its time evolution $\mathcal{U}(t)$ for $0\le t\le T$, not just $\mathcal{U}(T)$.

\subsection{FTI with trivial Floquet operator}
It is convenient to start with an FTI with a trivial Floquet operator satisfying $\mathcal{U}(T)=1$. In this case, the infinitesimally thin quasi-energy band is located at $\epsilon=0$. There is an energy gap extending from $\epsilon=0^+$ to $\pi/T$, which is equivalent to $-\pi/T$, then back to $\epsilon=0^-$. When periodic boundary condition in real space is imposed, $\mathcal{U}(\vk,t)$ is periodic in $k_x$ and $k_y$. The condition $\mathcal{U}(0)=\mathcal{U}(T)=1$ makes $\mathcal{U}(\vk,t)$ periodic in $t$ as well. Therefore, $\mathcal{U}(\vk,t)$ defines a mapping from a 3D torus $T^3$ to the unitary group $U(N)$. This mapping is classified by the homotopy group $\pi_3(U(N))=Z$. Therefore, it can be characterized by the winding number $W(\mathcal{U})$, given by
\be\label{eq:WU}
\frac{1}{8\pi^2}\int dtdk_xdk_y \mbox{Tr}\Big(\mathcal{U}^{-1}\p_t \mathcal{U}
\Big[\mathcal{U}^{-1}\p_{k_x}\mathcal{U},\,\mathcal{U}^{-1}\p_{k_y}\mathcal{U}\Big]\Big).
\ee
It can be shown that the winding number is also equal to the number of edge modes inside the energy gap \cite{Rudner}.

For a generic time-dependent Hamiltonian, the time evolution operator usually can not be obtained exactly. One has to use a perturbation expansion to approximately evaluate $\mathcal{U}(t)$. As an example, we consider a 2D two-band model described by the Hamiltonian
\begin{equation}\label{eq:H0}
H=\sum_{i=x,y,z} n_i\sigma_i.
\end{equation}
Here $n_i$ with $i=x,y,z$ are functions of $k_x$ and $k_y$ satisfying $\sum_i n_i^2=1$, and $\sigma_i$ are the Pauli matrices. In the following we will use the Einstein convention and sum over repeated indexes. This Hamiltonian is independent of time, so its time evolution operator is
\be
\mathcal{U}(\vk,t)=\exp[-i H(\vk)t]=\cos t-i\sin t(n_a\sigma_a).
\ee
The driving period is $T=2\pi$ in this case, and $\mathcal{U}(2\pi)=1$ leads to a trivial Floquet operator.
The inverse of $\mathcal{U}(t)$ is given by $\mathcal{U}^{-1}=\cos t+i\sin t(n_a\sigma_a)$. Then,
\be
&&\mathcal{U}^{-1}\p_t\mathcal{U}=-in_a\sigma_a, \\
&&\Big[\mathcal{U}^{-1}\p_{k_x}\mathcal{U},\,\mathcal{U}^{-1}\p_{k_y}\mathcal{U}\Big]
=2\sin^2t\epsilon_{abc}\sigma_a(\p_{k_x}n_b)(\p_{k_y}n_c). \nonumber \\
\ee
Making use of the above results, we find the winding number $W(\mathcal{U})=2w(n)$, where
\be
&&w(n)=\frac{1}{4\pi}\int dk_xdk_y \epsilon_{abc}n_a(\p_{k_x}n_b)(\p_{k_y}n_c).
\ee
Here $n_i$ defines a mapping from a 2D torus to a 2D sphere. The above result shows that $W(\mathcal{U})$ is twice the winding number $w(n)$ of the mapping defined by $n_i$.

\subsection{FTI with nontrivial Floquet operator}\label{subsec:Nontrivial}
For the general case with $\mathcal{U}(\vk,T)\neq1$, the quasi-energy band will be more complicated. Suppose one want to consider the winding number at certain quasi-energy $E_0$, one can smoothly deform $\mathcal{U}(\vk,t)$ without closing the energy gap to a new time evolution operator $\mathcal{U}_{E_0}(\vk,t)$ such that $\mathcal{U}_{E_0}(\vk,T)=1$. The new time evolution operator can be constructed as follows.
\be
\mathcal{U}_{E_0}(\vk,t)=\left\{
\begin{array}{ll}
	\mathcal{U}(\vk,2t), & 0<t<T/2, \\
	\exp\Big[-iH_{\textrm{eff}}(2T-2t)\Big], & T/2<t<T,
\end{array}
\right.
\ee
where $H_{\textrm{eff}}=\frac{i}{T}\ln \mathcal{U}(\vk,T)$ is the effective Hamiltonian corresponding to the Floquet operator. One can show that the second-stage evolution reverses the first-stage evolution, therefore $\mathcal{U}_{E_0}(\vk,T)=1$. The winding number can then be found by Eq.~\eqref{eq:WU} for the case with a trivial Floquet operator.

To smoothly connect this new time evolution $\mathcal{U}_{E_0}(\vk,t)$ to the old one, we can consider the following family of time-evolution operators with an additional parameter $0\le s \le 1$. (Let $T_h =T/(1+s)$.)
\be
\mathcal{U}_s(\vk,t)&=&\left\{
\begin{array}{ll}
	\mathcal{U}(\vk,(1+s)t), & 0<t<T_h, \\
	\exp\Big[-iH_{\textrm{eff}}(2T-(1+s)t)\Big], & T_h<t<T.
\end{array}
\right. \nonumber \\
& &\label{eq-Us}
\ee
It can be shown that
\be
\mathcal{U}_{s=0}(\vk,t)=\mathcal{U}(\vk,t),\quad \mathcal{U}_{s=1}(\vk,t)=\mathcal{U}_{E_0}(\vk,t),
\ee
and the energy gap is not closed during the deformation. Since the winding number is a topological invariant, its value does not change under a smooth deformation. Therefore, the winding number and the number of edge modes of a general case with a nontrivial Floquet operator can be informed from the winding number $W(\mathcal{U}_{E_0})$.

\subsection{Examples}
As a concrete example, we take the Hamiltonian~\eqref{eq:H0} with $n_i=\mathcal{R}_i/|\mathcal{R}_i|$, where
\be\label{eq:R}
(\mathcal{R}_x,\mathcal{R}_y,\mathcal{R}_z)=(\sin k_x,\sin k_y,m+\cos k_x+\cos k_y).
\ee
It has been shown, for $0<m<2$ and $-2<m<0$, $w(n)=1$ and $w(n)=-1$, respectively~\cite{Qi2}. The corresponding 2D FTIs with the same parameters thus have the winding numbers $W(\mathcal{U})=2$ and $W(\mathcal{U})=-2$, respectively.

However, the aforementioned simple model does provide an example with a unit winding number. To construct a model with $W(\mathcal{U})=1$, we consider the following piecewise time-independent Hamiltonian
\be
H_1=\left\{
    \begin{array}{ll}
      n_i\sigma_i, & 0<t<\pi, \\
      1, & \pi<t<2\pi,
    \end{array}
  \right.
\label{H1}
\ee
where $n_i$ is given by Eq.~\eqref{eq:R}.
The corresponding time-evolution operator is
\be
\mathcal{U}_1=\left\{
    \begin{array}{ll}
      \cos t-i\sin t\,(n_i\sigma_i), & 0<t<\pi ;\\
      \exp(-i\,t), & \pi<t<2\pi.
    \end{array}
  \right.
\ee
It can be shown that at the end of the first stage of time evolution, $\mathcal{U}_1(\pi)=-1$. In the second stage, the Hamiltonian is a constant, which is chosen to bring in an extra minus sign for the time-evolution operator. Therefore, for $H_1$ we find that its time evolution still satisfies $\mathcal{U}_1(2\pi)=1$. For this model, the integration in the computation of the winding number only receives a finite contribution from half of the time. Thus, we find the winding number to be $W(\mathcal{U}_1)=1$ for $0<m<2$ and $W(\mathcal{U}_1)=-1$ for $-2<m<0$.

One may generalize the example to a 2D FTI with an explicit time dependence, given by
\be
&&H=A\Big[\sin k_x\sigma_x+\sin k_y\sigma_y\nonumber\\
&&+(m+w\cos t+\cos k_x+\cos k_y)\sigma_z\Big],
\ee
where $0\le t<2\pi$, $A$ is an overall multiplicative constant, and $w$ is the amplitude of the periodic driving force. For this model, the Floquet operator $\mathcal{U}(2\pi)\neq 1$ in general. Nevertheless, one can follow the method presented in Sec.~\ref{subsec:Nontrivial} and deform the time-evolution operator of the above Hamiltonian to one with a trivial Floquet operator. That way its winding number can be inferred from the deformed trivial Floquet operator. For example, we take $A=0.1$, $m=-0.5$, and $w=0.5$ and find that the quasi-energy spectrum has a gap at $\epsilon=0$. The winding number corresponding to the gap can be obtained from the deformed time-evolution operator, and its value is one. We will show that there is an edge mode located around $\epsilon=0$ in Section \ref{sec-edge}, compatible with the counting from the winding number.

\section{3D FTI with $Z_2$ index}\label{sec-3d}
Next, we extend the 2D two-band model discussed above with one more spatial dimension. As a consequence, the time-evolution operator $\mathcal{U}(\vk,t)$ defines a mapping from a 4D torus $T^4$ to the group $U(2)$. If we ignore the non-trivial cycles on $T^4$, the topological classification of this mapping is given by the homotopy group~\cite{Puttmann03} $\pi_4(U(2))=\pi_4(S^3)=Z_2$. Here the replacement of $T^4$ by $S^4$ in the homotopy group follows Ref.~\cite{Kennedy15}. To understand this homotopy group usually requires tools from algebraic topology such as spectral sequences~\cite{Bott_book}. Here, we use an alternative method to compute the $Z_2$ index shown in Refs.~\cite{Witten,Witten1}.

We define the following 3D two-band model by introducing a similarity transformation to a periodic, time dependent 2D two-band Hamiltonian $H_{2d}$:
\be\label{eq:SimTran}
H_2=U_z^{-1}H_{2d}U_z,~~ U_z=\exp(-i\frac{k_z}{2}\sigma_z).
\ee
Note that the unitary operator $U_z$ is not periodic in $k_z$ (i.e., $U_z(k_z=0)\neq U_z(k_z=2\pi)$), but $H_2$ is. Moreover, $H_{2d}$ may be a stack of 2D FTI with the same winding number, and we will give an examples later. If the time-evolution operator of $H_{2d}$ is $\mathcal{U}_{2d}(k_x,k_y,t)$, the time-evolution operator of $H_2$ is given by
\be
\mathcal{U}_2(k_x,k_y,k_z,t)=U_z^{-1}\mathcal{U}_{2d}(k_x,k_y,t)U_z,
\ee
which defines a mapping from $T^3\times S^1$ to $U(2)$.

\subsection{3D FTI with trivial time evolution operator}
We start with the case with a trivial Floquet evolution operator $\mathcal{U}_{2d}(T)=1$. To derive the $Z_2$ index of the 3D two-band FTI, we embed the $U(2)$ time-evolution operator into the $U(3)$ group as follows. Let
\be
V_1=\left(
              \begin{array}{cc}
                \mathcal{U}_{2d} & 0 \\
                0 & 1 \\
              \end{array}
            \right).
\ee
We introduce the similarity transformation
\be
V_2=\left(
      \begin{array}{ccc}
        e^{ik_z/2} & 0 & 0 \\
        0 & e^{-ik_z/2} & 0 \\
        0 & 0 & 1 \\
      \end{array}
    \right)
V_1\left(
      \begin{array}{ccc}
        e^{-ik_z/2} & 0 & 0 \\
        0 & e^{ik_z/2} & 0 \\
        0 & 0 & 1 \\
      \end{array}
    \right).
\ee
Without changing the embedded $U(2)$ part, we can also rewrite the above transformation as
\be
V_2&=&\left(
      \begin{array}{ccc}
        1 & 0 & 0 \\
        0 & e^{-ik_z} & 0 \\
        0 & 0 & e^{ik_z} \\
      \end{array}
    \right)
V_1\left(
      \begin{array}{ccc}
        1 & 0 & 0 \\
        0 & e^{ik_z} & 0 \\
        0 & 0 & e^{-ik_z} \\
      \end{array}
    \right) \nonumber \\
    &\equiv& M^{-1} V_1 M.
\label{eq-V2}
\ee

In order to define the topological index, one can treat the $S^1$ of the periodic $k_z$ as the boundary of a 2D unit disc and extend the definition of the similarity transformation from the boundary into the bulk by replacing $M$ in Eq.~\eqref{eq-V2} by
\be
R(k_z,r)=\left(
     \begin{array}{ccc}
       1 & 0 & 0 \\
       0 & re^{ik_z} & \sqrt{1-r^2} \\
       0 & -\sqrt{1-r^2} & re^{-ik_z} \\
     \end{array}
   \right).
   \label{eq-Rkz}
\ee
Here we introduce a new parameter $0\le r \le 1$, which is the radius of the 2D disc. When $r=1$, we get back to the similarity transformation of Eq.~(\ref{eq-V2}). Making use of $R(k_z,r)$ to transform $V_1$, we obtained the following $U(3)$ time-evolution operator
\be
g(k_x,k_y,t,k_z,r)=R(k_z,r)^{-1}V_1 R(k_z,r).
\label{RVR}
\ee
To be precise, the above similarity transformation actually defines a mapping from $T^3\times D^2$ to the coset space $U(3)/U(2)$. The is because the $U(2)$ transformation on the boundary, Eq.~\eqref{eq-V2}, is fixed, but the $U(3)$ transformation in the bulk, Eq.~\eqref{RVR}, may be varied.

The $Z_2$ topological index of the 3D two-band FTI is given by $\mathcal{I}=\exp[i\Gamma(g)]$, where~\cite{Witten1}
\be
\Gamma(g)=\frac{-i}{240\pi^2}\int_{T^3\times D^2}\mbox{Tr}(g^{-1}dg)^5.
\label{ga}
\ee
Here we use the language of differential form to simplify the notations and omit the exterior product symbol $\bigwedge$ between differential forms. Eq.~(\ref{ga}) actually defines the winding number of the mapping $ T^3\times D^2\to U(3)/U(2)$. $(\Gamma(g)/\pi)$ is also known as the Wess-Zumino amplitude, which has been discussed in the context of topological insulators~\cite{Gawedzki17}. 

We emphasize that the Wess-Zumino amplitude is a general topological index with its value taken from $Z_2$, so the way one constructs a 3D model does not affect its value associated with the 3D model. For a given 3D two-band Hamiltonian $H_{3d}(k_x,k_y,k_z)$, one can calculate the time-evolution operator $\mathcal{U}(k_x,k_y,k_z,t)$, which is a 2 by 2 matrix. It can be embedded into a 3 by 3 matrix $V$ by
\be
V=\left(
              \begin{array}{cc}
                \mathcal{U} & 0 \\
                0 & 1 \\
              \end{array}
            \right).
\ee
Then, one finds a smooth one-parameter extension of $V$ denoted by $g(r)$ with $0\le r\le 1$, such that $g(r=1)=V$, similar to the procedure shown in Eq.~\eqref{RVR}. In terms of $g$, the Wess-Zumino amplitude is again given by Eq.(\ref{ga}), apart from a factor of $\pi$. Therefore, for a given 3D two-band FTI, it is possible to carry out the calculation of $\Gamma$ numerically by identifying a smooth extension $g(r)$. Although the extension of $V$ to $g$ is not necessarily unique, the parity of $(\Gamma/\pi)$ does not depend on any specific choice of $g$, as shown in Ref.~\cite{Witten1}

Here, we demonstrate a particular construction of 3D two-band FTI shown in Eq.~(\ref{eq:SimTran}), which can be thought of as a straightforward way because it starts from a 2D two-band model with a $Z$-topological index. For the 3D two-band model of Eq.~(\ref{eq:SimTran}), we will show that $\Gamma(g)$ can also be computed analytically by a relation to the winding number $W(\mathcal{U})$ of the corresponding 2D model shown in Eq.~\eqref{eq:WU}.
The key step is to make use of the so-called Polyakov-Wiegmann identity~\cite{Hatsugai,Monaco17}
\be
\Gamma(gh)&=&\Gamma(g)+\Gamma(h)+\Delta\Gamma(g,h)\label{gh},\\
\Delta\Gamma(g,h)&=&\frac{i}{48\pi^2}\int_{T^3\times S^1}\mbox{Tr}\Big[(g^{-1}dg)^3dhh^{-1} \nonumber \\
& &+g^{-1}dg(dhh^{-1})^3+\frac12(g^{-1}dg dhh^{-1})^2\Big].\nonumber
\ee
The derivation is summarized in the Appendix.
One obtains, from Eq.~\eqref{RVR},
\be
&&\Gamma(g)=\Gamma(R^{-1})+\Gamma(V R)+\Delta\Gamma(R^{-1},V R),\\
&&\Gamma(V R)=\Gamma(V)+\Gamma(R)+\Delta\Gamma(V,R).
\ee
Since $R$ from Eq.~\eqref{eq-Rkz} only depends on the two variables $(k_z,r)$ and $V_1$ only depends on the three variables $(k_x,\,k_y,\,t)$, the anti-symmetric property of the wedge product gives rise to $(R^{-1}dR)^5=(V^{-1}dV)^5=0$, which also means that $\Gamma(R^{-1})=\Gamma(R)=\Gamma(V)=0$. Therefore, one obtains
\be
\Gamma(g)=\Delta\Gamma(R^{-1},V R)+\Delta\Gamma(V, R).
\ee
Following a similar argument, $(dR R^{-1})^3=0$. Furthermore, one can show that
\be
\int_{T^3\times S^1}\mbox{Tr}\Big(V^{-1}d VdRR^{-1}\Big)^2=0.
\ee
The reason is that the integrand is proportional to $dr$, but the boundary of $T^3\times D^2$
has a fixed value of $r$, causing the above integrals to vanish. Therefore, we obtain
\be
\Delta\Gamma(V,R)=\frac{i}{48\pi^2}\int_{T^3\times S^1}\mbox{Tr}[(V^{-1}dV)^3dRR^{-1}].
\ee
Similarly, one can show that
\be
\Delta\Gamma(R^{-1},VR)&=&-\frac{i}{48\pi^2}\int_{T^3\times S^1}\mbox{Tr}[dRR^{-1}(dVV^{-1})^3]. \nonumber \\
& &
\ee
Collecting the above results, we find
\be\label{eq:GammaW}
\Gamma(g)&=&\frac{-i}{48\pi^2}\int_{T^3\times S^1}
\mbox{Tr}\Big(dRR^{-1}\Big[(dVV^{-1})^3+(V^{-1}dV)^3\Big]\Big)\nonumber\\
&=&\frac{1}{24\pi}\int_{T^3}\mbox{Tr}(\mathcal{U}_{2d}^{-1}d\mathcal{U}_{2d})^3 \nonumber \\
&=&\pi W(\mathcal{U}_{2d}).
\ee

\subsection{Relations between topological quantities}\label{subsec:relations}
There is a connection between the winding number defined by Eq.~(\ref{ga}) and the elements of $\pi_4(U(2))=Z_2$. We provide a heuristic argument as follows. One can treat the $U(3)$ group as a $U(2)$ principal bundle over the base manifold $U(3)/U(2)$, then there exists a long exact homotopy sequence for fiber bundles~\cite{Bott_book}
\be
& &\cdots \to \pi_{k+1}(U(3)/U(2)) \to \pi_k(U(2)) \to \pi_k(U(3)) \nonumber \\
& &\to \pi_{k}(U(3)/U(2)) \to \pi_{k-1}(U(2))\to\cdots .
\ee
Here the exact sequence means that the image of each mapping in the sequence equals the kernel of the next mapping.
From the known fact~\cite{Nakahara_book} $\pi_4(U(3))=0$, the above sequence is truncated as
\be
& &\cdots \to \pi_5(U(3))\to \pi_{5}(U(3)/U(2)) \to \pi_{4}(U(2))\to 0. \nonumber \\
& &
\ee
Moreover, it is known that~\cite{Puttmann03,Nakahara_book} $\pi_5(U(3))=Z$ and $\pi_4(U(2))=Z_2$.

The coset space $U(3)/U(2)$ is equivalent to the five-dimensional sphere (see Section 17.5 of Ref.~\cite{Nair}), which gives rise to $\pi_{5}(U(3)/U(2))=Z$. Thus, the above exact sequence translates to
\be
\cdots\to Z\xrightarrow{\times 2} Z\to Z_2\to0.
\ee
A more detailed analysis shows that the labeled arrow in the above sequence represents the multiplication by $2$. The exactness of the above sequence requires that the non-trivial element of $Z_2$ is the image of the odd numbers of the previous $Z$. Since the winding number from Eq.~(\ref{ga}) corresponds to $\pi_5(SU(3)/SU(2))$, the parity will decide where it will be mapped to $\pi_4(U(2))$. It can be shown that the odd winding numbers of $\pi_{5}(SU(3)/SU(2))$ map to the non-trivial element of $\pi_4(U(2))$, and the even winding numbers map to the trivial one.

For the 2D two-band FTI, the winding number $W(\mathcal{U})$ has been shown to relate to the Chern number computed from the eigenstates of the Floquet operator~\cite{Rudner}. This raises the question if some similar relation also exists for the 3D two-band FTI. The answer is affirmative and can be understood as follows. The eigenstates of the Floquet operator of a 3D two-band FTI defines a 3D unit vector $s_i=\bra{\psi_n}\sigma_i\ket{\psi_n}$, which can be thought of as a mapping from $T^3$ to $S^2$. After we ignore the nontrivial cycles~\cite{Kennedy15} of $T^3$, the mapping of the eigenstates can be classified by the Hopf index of $\pi_3(S^2)=Z$. One can increase the dimension of an $n$-sphere by suspension (or smash product) with $S^1$ as $S^{n+1}=S^1\wedge S^n$ (not to be confused with the symbol of exterior product). The technique leads to a mapping $\pi_3(S^2)\to\pi_4(S^3)$. As a consequence, the odd Hopf indices map to the nontrivial element of $\pi_4(S^3)=Z_2$ and the even Hopf indices map to the trivial one (see Section 11.15 of Ref.~\cite{S-T-Hu}).

\subsection{FTI with nontrivial Floquet operator}
In the above discussion, we only construct the 3D two-band FTI from a 2D FTI (or a stack of 2D FTI) with trivial Floquet operator $\mathcal{U}_{2d}(\vk,T)=1$. Similar procedure can also be applied to a more general 2D FTI (or a stack of 2D FTI) with $\mathcal{U}_{2d}(\vk,T)\neq 1$. Following the steps in Section \ref{subsec:Nontrivial}, we introduce a new time evolution operator for a given quasi-energy gap $E_0$ as
\be
\mathcal{U}_{E_0}(\vk,t)=\left\{
\begin{array}{ll}
	\mathcal{U}_{2d}(\vk,2t), & 0<t<T/2, \\
	\exp\Big[-iH_{2d,\textrm{eff}}(2T-2t)\Big], & T/2<t<T,
\end{array}\nonumber
\right.
\ee
where $H_{2d,\textrm{eff}}=\frac{i}{T}\ln \mathcal{U}_{2d}(\vk,T)$. By definition, $\mathcal{U}_{E_0}(T)=1$ and we can apply the formulas for the case with a trivial Floquet operator. $\mathcal{U}_{E_0}$ is smoothly connected to $\mathcal{U}_{2d}$ without closing the quasi-energy gap as shown Eq.~(\ref{eq-Us}).

The calculation of the $Z_2$ index based on the new time evolution operator $\mathcal{U}_{E_0}$ can be performed by the procedure described in the previous subsection: We embed the above $U(2)$ operator into a $U(3)$ operator by a similarity transformation
\be
g(\vk,t,r)=R(k_z,r)^{-1}\left(
                          \begin{array}{cc}
                            \mathcal{U}_{E_0} & 0 \\
                            0 & 1
                          \end{array}
                        \right)
R(k_z,r),
\ee
where $R(k_z,r)$ is given by Eq.~(\ref{eq-Rkz}). The $Z_2$ index $\Gamma(g)$ is then computed by Eq.~(\ref{ga}). Since a smooth deformation between the cases with and without a nontrivial Floquet operator does not alter the quantized topological invariant, the general case will have the same $Z_2$ index obtained from its counterpart with a trivial Floquet operator.

\subsection{Examples}
As an example, we extend the 2D FTI model of Eq.~(\ref{H1}) with piecewise-constant time-dependence to a 3D two-band FTI by using Eq.~\eqref{eq:SimTran} and the procedure following it. Our numerical calculation verifies that $\Gamma(g)=\pm\pi$ for $0<m<2$ and $-2<m<0$, respectively. For both parameter regimes, we have $\mathcal{I}=-1$, so they correspond to the topologically non-trivial mapping. On the other hand, for $m>2$ or $m<-2$, we find that $\Gamma(g)=0$ and $\mathcal{I}=1$, which corresponds to the trivial case. Since $W(\mathcal{U}_1)=\pm 1$ for the model of Eq.~\eqref{H1} with $0<m<2$ and $-2<m<0$, respectively, one can infer that $\Gamma(g)=\pm\pi$ from Eq.~\eqref{eq:GammaW} for the corresponding 3D two-band FTI. In both parameter regimes, we have $\mathcal{I}=-1$, which agrees with the numerical result. In the next section, we will show more general models of 3D two-band FTI showing explicit time dependence.

\begin{figure*}[th]
\centerline{\includegraphics[width=0.8\textwidth]{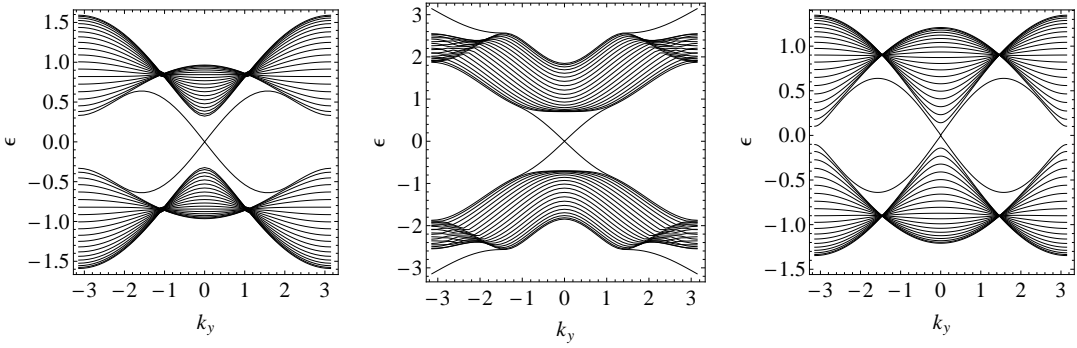}}
\caption{Left panel: Quasi-energy spectrum as a function of $k_y$ for the 2D two-band FTI model of Eq.~(\ref{H-2d}) with one edge mode located around $\epsilon=0$. Middle panel: Same model as the left panel with different parameters. There are one edge mode located around $\epsilon=0$ and another one located at $\epsilon=\pm\pi$. Right panel: Quasi-energy spectrum as a function of $k_y$ for the 3D two-band FTI model of Eq.~(\ref{eq:BlochH3d}) with $k_z=0.3$.}
\label{fig-x}
\end{figure*}

\begin{figure*}[th]
	\centerline{\includegraphics[width=0.8\textwidth]{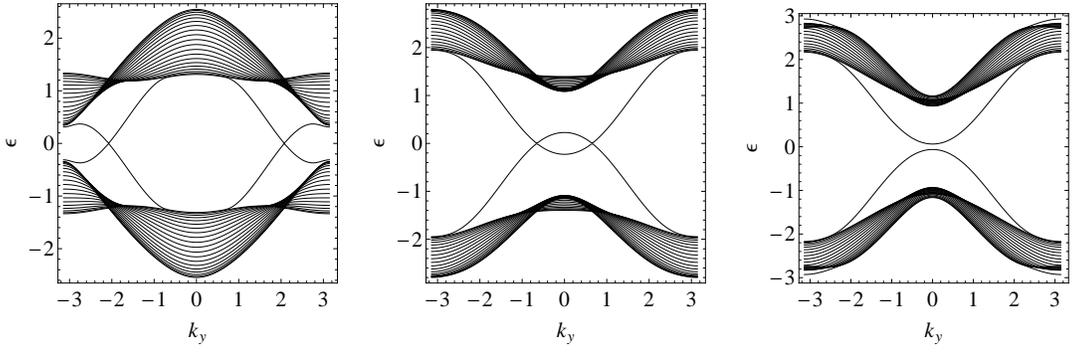}}
	\caption{The quasi-energy spectrum as a function of $k_y$ for the 3D two-band FTI of Eq.~(\ref{H3dz}) with open boundary condition in the $z$ direction. The panels from left to right correspond to $k_x=0.3,\,1.9,\,2.2$, respectively.}
	\label{fig-z}
\end{figure*}

\section{Edge modes}\label{sec-edge}
The topology of a system can be characterized by the topological index calculated from a manifold with periodic boundary condition, as we did in the previous sections. It can also be detected by studying the same Hamiltonian with open boundary condition, as we will analyze in this section. We will also discuss the relation between the number of edge modes and the $Z_2$ index of the 3D two-band FTI.

\subsection{2D two-band FTI}
We start with a 2D two-band FTI with periodic boundary condition described by the Bloch Hamiltonian
\be\label{eq:H2d}
&&H^{2d}=A\Big[\sin k_x\sigma_x+\sin k_y\sigma_y\nonumber\\
&&+(m+w\cos t+\cos k_x+\cos k_y)\sigma_z\Big].
\ee
Here $A$, $m$, and $w$ are parameters and the driving period is $T=2\pi$. Since the Floquet operator of Eq.~\eqref{eq:H2d} shows $\mathcal{U}(T)\neq 1$, we cannot directly apply the winding number formula. As discussed in Section \ref{sec-2d}, one can always smoothly deform the time-evolution operator to obtain a new one with $\mathcal{U}(T)=1$. Then, the winding number formula can be used to find the winding number at a selected value of quasi-energy for the new case with a trivial Floquet operator. Since topological invariants do not change under smooth deformation, the winding number remains the same for the original case before the deformation. For example, the winding number of the model~\eqref{eq:H2d} at $\epsilon=0$ is $W_\epsilon=1$.

The Hamiltonian corresponds to the same model~\eqref{eq:H2d} but with open boundary condition along the $x$-axis is
\be
&&H^{2d}_x=A\Big[\frac{\sigma_z-i\sigma_x}{2}\otimes h_0+h.c.\nonumber\\
&&+[\sin k_y\sigma_y+(m+w\cos t+\cos k_y)\sigma_z]\otimes I_0\Big],
\label{H-2d}
\ee
where $h_0=\delta_{i+1,j}$ and $I_0=\delta_{ij}$ are $N_x\times N_x$ matrices with $i,j=1,\cdots,N_x$. Here $N_x$ is the number of sites along the $x$ axis. The left panel of Figure \ref{fig-x} shows the quasi-energy spectrum as a function of $k_y$ with $N_x=20$, $A=0.1$, $m=-0.5$ and $w=0.5$. There is one chiral edge mode at each end point of the $x$ direction, consistent with the winding number $W_\epsilon=1$ at $\epsilon=0$. In contrast, the middle panel of Figure \ref{fig-x} shows the spectrum with $A=0.2$, $m=-1.5$, and $w=2.5$. In addition to one edge mode at each end point around $\epsilon=0$, there is another edge mode at each end point around $\epsilon=\pm\pi$ in this case. The extra edge mode is the so-called anomalous edge mode~\cite{Rudner}, an interesting feature of the periodic driven system.

\subsection{3D two-band FTI}
We next construct a 3D two-band FTI related to the 2D two band FTI ~\eqref{eq:H2d}. The Bloch Hamiltonian of the 3D system, $H^{3d}$, with periodic boundary condition can be obtained by a similarity transformation of a two-band Hamiltonian $H^{st}$. Explicitly,
\be\label{eq:BlochH3d}
&&H^{st}=A\Big[\sin k_x\sigma_x+\sin k_y\sigma_y\nonumber\\
&&+(m+w\cos t+m'\cos k_z+\cos k_x+\cos k_y)\sigma_z\Big], \nonumber \\
&&H^{3d}=U_z^{-1}H^{st}U_z .
\ee
Here $U_z=\exp(i\sigma_z k_z/2)$ and we introduce an additional momentum component $k_z$ to the model. By comparing $H^{st}$ with Eq.~(\ref{eq:H2d}), one can see that the parameter $m$ has been replaced by $m+m'\cos k_z$, which depends on the momentum of the new direction. Since the 2D FTI Eq.~(\ref{eq:H2d}) has quantized winding numbers, a small change of the value of $m$ does not change the winding number if the gap remains open. Therefore, as long as the amplitude $m'$ of the additional modulation in $H^{st}$ is not too large, each slice in the momentum space with fixed $k_z$ can be shown to be a 2D FTI with the same winding number. Thus, $H^{st}$ may be considered as a stack of 2D FTI with very weak $k_z$ dependence in the absence of the similarity transformation. The $Z_2$ index of the 3D two-band FTI~\eqref{eq:BlochH3d} at a selected quasi-energy can be obtained numerically following the procedure outlined in Sec.~\ref{sec-3d}.

In general, one can write the Hamiltonian in real space. For example, the above model can also be expressed by the equivalent Hamiltonian in the second quantized form as
\be
&&H^{3d}=\sum_n A\Big[-\dc_n\frac{i\sigma_+}{2}c_{n+\hat{x}-\hat{z}}
+\dc_n\frac{i\sigma_+}{2}c_{n-\hat{x}-\hat{z}}\nonumber\\
&&-\dc_n\frac{\sigma_+}{2}c_{n+\hat{y}-\hat{z}}+\dc_n\frac{\sigma_+}{2}c_{n-\hat{y}-\hat{z}}
+h.c.\nonumber\\
&&+\dc_n\frac{\sigma_z}{2}c_{n+\hat{x}}+\dc_n\frac{\sigma_z}{2}c_{n+\hat{y}}
+(m+w\cos t)\dc_n\sigma_z c_n
\Big].
\ee
Here $\dc_n=(\dc_{n_1},\dc_{n_2})$ denote the creation operators of two fermions (due to the two bands) on site $n=(n_x,n_y,n_z)$ with $n_x,n_y,n_z\in Z$. $\hat{x}$ is the unit vector along the $x$ direction, etc. $\sigma_+=(\sigma_x+i\sigma_y)/2$. In real space, one can see that the model contains nearest and next-nearest neighbor hopping terms.

To study the edge modes of the 3D FTI model~\eqref{eq:BlochH3d}, we begin by assuming open boundary condition along the $x$ axis. The Hamiltonian then becomes
\be
&&H^{3d}_x=A\Big[U_z^{-1}\dfrac{\sigma_z-i\sigma_x}{2}U_z\otimes h_{0}+h.c. +U_z^{-1}[\sin k_y\sigma_y\nonumber\\
&&+(m+w\cos t+m'\cos k_z+\cos k_y)\sigma_z] U_z\otimes I_0\Big].
\label{H-3d}
\ee
For the 3D two-band FTI, the quasi-energy spectrum depends on $k_y$ and $k_z$ and forms a 2D curved surface. Here we only show the curves at fixed values of $k_z$ by plotting the quasi-energy spectrum as a function of $k_y$ in the right panel of Fig.~\ref{fig-x} with $k_z=0.3$, $A=0.1$, $m=-0.5$, $m'=0.4$, and $w=0.5$. The spectrum of the 3D model is very similar to the 2D FTI in the left panel of Fig.~\ref{fig-x}. This is because $H^{3d}_x$ is a similar transformation of $H^{st}$ with open boundary condition, and the parameter $m+m'\cos k_z$ does not close the gap of Eq.~(\ref{H-2d}). Therefore, the quasi-energy spectra of $H^{3d}_x$ and $H^{2d}_x$ are similar. There is an edge mode at each end along the $x$ axis for the 3D two-band FTI. Incidentally, one may consider $H^{st}$ with open boundary condition for a stack of 2D FTIs with the same winding number, and there is one edge mode at each end weakly dependent on $k_z$.

\begin{figure}[th]
\centerline{\includegraphics[width=\columnwidth]{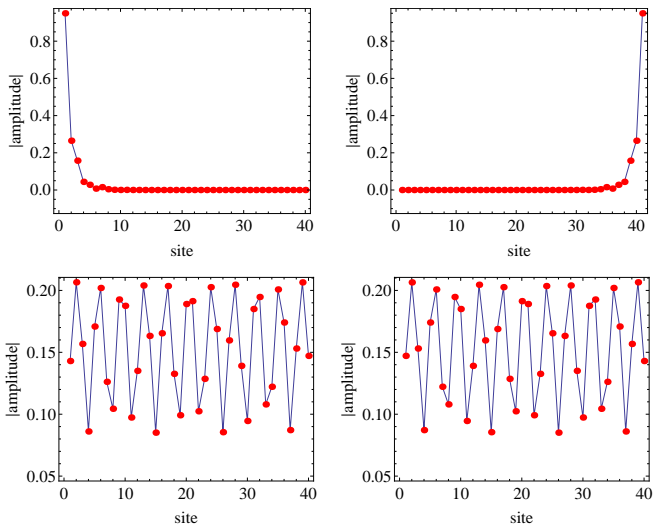}}
\caption{Top panels: Wavefunctions of the edge modes of Eq.~(\ref{H3dz}) located at the left and right boundaries, respectively. Lower left (right) panel: Wavefunction of one selected state from the top (bottom) bulk band.}
\label{wavefunc}
\end{figure}

To test the robustness of the edge modes, we compute the quasi-particle spectrum with a different open boundary condition by keeping the $x,\,y$ directions periodic but imposing open boundary condition in the $z$ direction. The Hamiltonian corresponding to Eq.~\eqref{eq:BlochH3d} in this configuration becomes
\be
&&H^{3d}_z=A\Big\{[(\sin k_x-i\sin k_y)\dfrac{\sigma_x-i\sigma_y}{2}+\frac{m'}{2}\sigma_z]\otimes h_{0}+h.c.\nonumber\\
&&+(m+w\cos t+\cos k_x+\cos k_y)\sigma_z\otimes I_0\Big\},
\label{H3dz}
\ee
Here $h_0$ and $I_0$ have the same structures except $N_z$ replaces $N_x$, and we assume there are $N_z$ sites in the $z$ direction. The quasi-energy spectrum of Eq.~(\ref{H3dz}) is shown in Figure \ref{fig-z} with $N_z=20$, $A=0.2$, $m=-0.5$, $w=1.5$, and $m'=0.5$. From the left to right panels of Figure \ref{fig-z}, the spectrum as a function of $k_y$ is plotted for $k_x=0.3,\,1.9,\,2.2$, respectively. From the left to right, one can see that the distance between the two crossings of the edge modes gradually shrinks to zero, and then the quasi-energies of the edge modes no longer intersect. From this observation, one can deduce that the edge modes actually have paraboloid-shaped dispersions, and there is only one edge mode for each end point along the $z$ direction. This is also consistent with the result we have found for the same model with open boundary along the $x$ direction shown in the right panel of Fig.~\ref{fig-x}. However, the emergence of the edge mode in the $z$ direction differentiate the 3D FTI from a stack of 2D FTI because a stack of 2D FTI does not support a topological edge mode along the direction of its stacking. Since the edge mode appears in the newly introduced boundary in the $z$ direction, which is also the stacking direction of $H^{st}$, the edge mode is intrinsic to the 3D system, not associated with the stacking of 2D FTI in $H^{st}$.

In order to better characterize the edge modes, we also sample the wavefunctions of typical modes and show them in Figure \ref{wavefunc} for the case with $k_x=0.3$ and $k_y=-1.6$. The top panels show the wavefunctions of the two edge modes corresponding to quasi-energy $\epsilon=\pm 0.48$, respectively. One can see that they are indeed localized to the two ends of the open boundary. The bottom panels shows the wavefunctions of two selected states, one from the top bulk band and one from the bottom bulk band, corresponding to the quasi-energy $\epsilon=\pm 1.51$, respectively. For the bulk states, one can see that they spread across the whole system. In Figure \ref{wavefunc}, we present the results for a larger lattice with $N_z=40$ sites in order to show more details of the wavefunctions.

\begin{figure}[th]
\centerline{\includegraphics[width=\columnwidth]{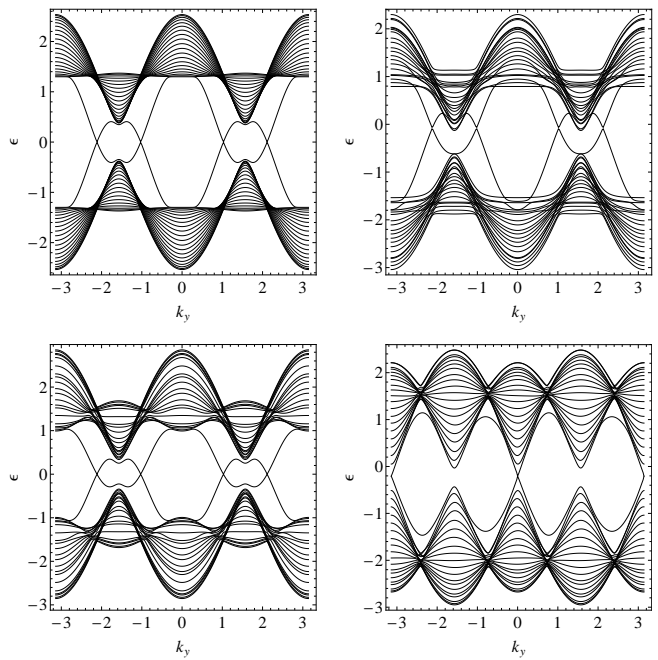}}
\caption{(Upper left) The quasi-energy spectrum as a function of $k_y$ from the 3D two-band FTI model of Eq.~(\ref{H2}) with open boundary condition in the $z$-direction and fixed $k_x=0.3$. (Upper right and lower left) The quasi-energy spectra as a function of $k_y$ and fixed $k_x=0.3$ of the 3D two-band FTI models with random onsite potentials of Eq.~(\ref{H2r}) and with random hopping coefficients of Eq.~(\ref{H2r1}), respectively. (Lower right) The quasi-energy spectrum of Eq.~(\ref{H2rx}) as a function of $k_y$ with random onsite potentials along the $x$ direction.} 
\label{fig-w2}
\end{figure}

\subsection{Robustness of the edge mode and bulk-boundary correspondence}
We have demonstrated that when the topological index $\Gamma=\pi$, the edge mode does exist in the band gap. This indicates a topologically non-trivial phase of the $Z_2$ classification. One may wonder what happens when $\Gamma=2\pi$.
The 3D two-band FTI with $\Gamma=2\pi$ can be obtained from the 2D two-band FTI by a similarity transformation. Thus, one would expect that there are two edge modes for the resulting 3D two-band FTI. To verify this assertion, we modify Eq.~(\ref{H3dz}) so that the modified Hamiltonian gives $\Gamma=2\pi$. Explicitly,
\be
&&H_{2\pi}=A\Big[[(\sin k_x-i\sin 2k_y)\dfrac{\sigma_x-i\sigma_y}{2}+\frac{m'}{2}\sigma_z]\otimes h_{0}+h.c.\nonumber\\
&&+(m+w\cos t+\cos k_x+\cos 2k_y)\sigma_z\otimes I_0\Big],
\label{H2}
\ee
Here $A=0.2$, $m=-0.5$, $w=0.5$, and $m'=0.5$.
On the upper left panel of Figure \ref{fig-w2}, we plot the quasi-energy spectrum of Eq.~(\ref{H2}) as a function of $k_y$ for fixed $k_x=0.3$ with $N_z=20$. One can see that there are four crossings of the edge-mode branches, indicating the shapes of two paraboloid-like edge-mode dispersions.

According to the $Z_2$ index, $\Gamma=2\pi$ is expected to be equivalent to the trivial case of $\Gamma=0$.
If the bulk-boundary correspondence for the 3D two-band FTI is analyzed in the way we did for the 2D two-band FTI, one expects that the two edge modes may not be stable under a $z$-coordinate dependent perturbation, and the two modes could annihilate each other. We will explain in the following this is not the case. Firstly, the edge modes are robust against a disorder potential in the $z$ direction. We show in the upper right and lower left panels of Figure \ref{fig-w2} the spectrum of the same system~\eqref{H2} with additional random onsite potentials or random hopping coefficients corresponding to the Hamiltonians
\be
&&H'_{2\pi}=H_{2\pi}+A\sigma_0\otimes I_r,\label{H2r}\\
&&H''_{2\pi}=H_{2\pi}+A\sigma_z\otimes h_r.
\label{H2r1}
\ee
Here $\sigma_0$ is the 2 by 2 identity matrix, $I_r=v_i\delta_{ij}$ and $h_r=v_i\delta_{i+1,j}$ for $i,j=1,\cdots,N_z$ with $v_i$ being a random variable uniformly distributed between $0$ and $0.5$. One can see that the bulk bands are broadened by the random potential or random hopping coefficient. The two edge modes, nevertheless, are still observable inside the band gap until the band gap closes due to the broadened bands. Therefore, no edge-mode annihilation occurs in the 3D two-band FTI.

Next, we consider the same model with open boundary condition in the $x$ direction and random onsite potentials along the $x$ direction. The Hamiltonian is given by
\be
&&H^{x}_{2\pi}=A\Big[U_z^{-1}\dfrac{\sigma_z-i\sigma_x}{2}U_z\otimes h_{0}+h.c.
+U_z^{-1}[\sin 2k_y\sigma_y\nonumber\\
&&+(m+w\cos t+m'\cos k_z+\cos 2k_y)\sigma_z] U_z\otimes I_0\Big]\nonumber\\
&&+A\sigma_0\otimes I_r.
\label{H2rx}
\ee
The quasi-energy spectrum is shown in the lower right panel of Figure~\ref{fig-w2}. Here we take $A=0.2$, $m=-0.5$, $w=0.5$, $m'=0.4$, $k_z=0.3$, and $I_r=v_i\delta_{i,j}$ with $v_i$ being a random variable uniformly distributed between $0$ and $0.3$. One can see that there are still two edge modes in the band gap. Therefore, the edge modes are stable against random onsite potentials along the $z$ or $x$ direction.

In general, the number of edge modes is determined by the $Z$-index, which is equivalent to its winding number, of the 2D two-band FTI on which the 3D two-band FTI is based. By constructing 3D two-band FTIs with $\Gamma=n\pi$, where $n$ is an integer, it can be shown that there are $n$ edge modes at each end point of the open boundary. From the explanation given in Sec.~\ref{subsec:relations}, all the odd (even) winding numbers are mapped to the nontrivial (trivial) element in $Z_2$. Thus, we propose the parity of the number of edge modes, not the number itself, reflects the $Z_2$ index of the 3D two-band FTI. We notice that there is also no simple bulk-boundary correspondence for the Hopf insulator~\cite{Duan} based on the homotopy group $\pi_3(S^2)=Z$. The 3D two-band FTI is based on the homotopy group $\pi_4(U(2))=Z_2=\pi_4(S^3)$, and both $\pi_3(S^2)$ and $\pi_4(S^3)$ are homotopy groups between manifolds of different dimensions. Therefore, the 3D two-band FTI may share similar features as the Hopf insulator and complicate the bulk-boundary correspondence.

There have been experimental probes for various FTIs~\cite{Wang13,Fleury16,Maczewsky17} and proposals for measuring the winding number of some 2D FTIs~\cite{Unal18}.
A direct measurement of the $Z_2$ index of the 3D two-band FTI can be challenging because one has to connect the index to some physical quantity. Measurements of the quasi-energy spectrum or edge modes of the 3D two-band FTI with open boundary condition, nevertheless, may reveal the similar spectrum and edge modes when compared to their corresponding 2D two-band FTI. The $Z_2$ index can also be inferred from the parity of the number of edge modes.

The study focuses on two-band models. For a multi-band system of real materials, if two quasi-energy bands are separated from the other bands with a large gap, the time-evolution operator may be written in an almost blocked form as
\be
\mathcal{U}(t)=\left(
  \begin{array}{cc}
    \mathcal{U}_{2\times2} & \cdots \\
    \vdots & \mathcal{U}_{(n-2)\times(n-2)} \\
  \end{array}
\right).
\ee
If the off-diagonal blocks have very small elements, one may concentrate on the part $\mathcal{U}_{2\times2}$ for the two almost isolated bands and calculate the topological index $\Gamma$ (or $\mathcal{I}$). However, the off-diagonal blocks and the other bands will act like an environment and blur the features we have discussed.

\section{conclusion}\label{sec:conclusion}
We showed the construction of a class of 3D two-band FTI with a $Z_2$ topological index based on the homotopy group $\pi_4(U(2))$. The Hamiltonian of the 3D two-band FTI is obtained from a similarity transformation of a corresponding 2D two-band FTI with a $Z$ topological index. As a consequence, the quasi-energy spectrum of the 3D FTI resembles that of the 2D FTI when open boundary condition is imposed. However, the parity of the total number of edge modes corresponds to the bulk index for the 3D two-band FTI. The edge modes of the 3D two-band FTI are robust against weak disorder which does not close the gap. Since conventional classification schemes of FTI do not apply to the 3D two-band FTI discussed here, this class of 3D FTI offers additional examples of the rich physics of Floquet topological systems.

\acknowledgments
Y. H. thanks useful discussions with Chang-yan Wang. Y.H. is supported by NSFC under Grants No. 11874272.

\appendix

\section{Derivation of Eq. (\ref{gh})}
Here we show a derivation of Eq.~(\ref{gh}).
Note that $(gh)^{-1}d(gh)=h^{-1}(x+y)h$ with $x=g^{-1}dg$ and $y=dh h^{-1}$. Then, we have
\be
\Gamma(gh)&=&\frac{-i}{240\pi^2}\int_{M}\mbox{Tr}(x+y)^5\nonumber\\
&=&\Gamma(g)+\Gamma(h)+\frac{-i}{48\pi^2}\int_M\mbox{Tr}(x^4y+y^4x+x^3y^2 \nonumber \\
& &+y^3x^2+x^2yxy+y^2xyx).
\ee
Making use of the facts that $dx=-x^2$, $dy=y^2$, $d(x^3)=-x^4$, $d(y^3)=y^4$, we find
\be
&&d\Big(x^3y+xy^3+\frac12xyxy\Big)\nonumber\\
&&=-x^4y-x^3y^2-x^2y^3-xy^4-x^2yxy-xy^2xy.
\ee
By using Stoke's theorem, we find
\be
\Gamma(gh)&=&\Gamma(g)+\Gamma(h)+\frac{i}{48\pi^2}\int_M\mbox{Tr}d\Big(x^3y+xy^3+\frac12xyxy\Big)\nonumber\\
&&
\ee
which gives Eq.~(\ref{gh}).


\begin{thebibliography}{42}%
	\makeatletter
	\providecommand \@ifxundefined [1]{%
		\@ifx{#1\undefined}
	}%
	\providecommand \@ifnum [1]{%
		\ifnum #1\expandafter \@firstoftwo
		\else \expandafter \@secondoftwo
		\fi
	}%
	\providecommand \@ifx [1]{%
		\ifx #1\expandafter \@firstoftwo
		\else \expandafter \@secondoftwo
		\fi
	}%
	\providecommand \natexlab [1]{#1}%
	\providecommand \enquote  [1]{``#1''}%
	\providecommand \bibnamefont  [1]{#1}%
	\providecommand \bibfnamefont [1]{#1}%
	\providecommand \citenamefont [1]{#1}%
	\providecommand \href@noop [0]{\@secondoftwo}%
	\providecommand \href [0]{\begingroup \@sanitize@url \@href}%
	\providecommand \@href[1]{\@@startlink{#1}\@@href}%
	\providecommand \@@href[1]{\endgroup#1\@@endlink}%
	\providecommand \@sanitize@url [0]{\catcode `\\12\catcode `\$12\catcode
		`\&12\catcode `\#12\catcode `\^12\catcode `\_12\catcode `\%12\relax}%
	\providecommand \@@startlink[1]{}%
	\providecommand \@@endlink[0]{}%
	\providecommand \url  [0]{\begingroup\@sanitize@url \@url }%
	\providecommand \@url [1]{\endgroup\@href {#1}{\urlprefix }}%
	\providecommand \urlprefix  [0]{URL }%
	\providecommand \Eprint [0]{\href }%
	\providecommand \doibase [0]{http://dx.doi.org/}%
	\providecommand \selectlanguage [0]{\@gobble}%
	\providecommand \bibinfo  [0]{\@secondoftwo}%
	\providecommand \bibfield  [0]{\@secondoftwo}%
	\providecommand \translation [1]{[#1]}%
	\providecommand \BibitemOpen [0]{}%
	\providecommand \bibitemStop [0]{}%
	\providecommand \bibitemNoStop [0]{.\EOS\space}%
	\providecommand \EOS [0]{\spacefactor3000\relax}%
	\providecommand \BibitemShut  [1]{\csname bibitem#1\endcsname}%
	\let\auto@bib@innerbib\@empty
	\bibitem [{\citenamefont {Hasan}\ and\ \citenamefont
		{Kane}(2010)}]{Kane_TIRev}%
	\BibitemOpen
	\bibfield  {author} {\bibinfo {author} {\bibfnamefont {M.~Z.}\ \bibnamefont
			{Hasan}}\ and\ \bibinfo {author} {\bibfnamefont {C.}~\bibnamefont {Kane}},\
	}\href@noop {} {\bibfield  {journal} {\bibinfo  {journal} {Rev. Mod. Phys.}\
		}\textbf {\bibinfo {volume} {82}},\ \bibinfo {pages} {3045} (\bibinfo {year}
		{2010})}\BibitemShut {NoStop}%
	\bibitem [{\citenamefont {Qi}\ and\ \citenamefont {Zhang}(2011)}]{Zhang_TIRev}%
	\BibitemOpen
	\bibfield  {author} {\bibinfo {author} {\bibfnamefont {X.-L.}\ \bibnamefont
			{Qi}}\ and\ \bibinfo {author} {\bibfnamefont {S.-C.}\ \bibnamefont {Zhang}},\
	}\href@noop {} {\bibfield  {journal} {\bibinfo  {journal} {Rev. Mod. Phys.}\
		}\textbf {\bibinfo {volume} {83}},\ \bibinfo {pages} {1057} (\bibinfo {year}
		{2011})}\BibitemShut {NoStop}%
	\bibitem [{\citenamefont {Shen}(2012)}]{ShenTI}%
	\BibitemOpen
	\bibfield  {author} {\bibinfo {author} {\bibfnamefont {S.-Q.}\ \bibnamefont
			{Shen}},\ }\href@noop {} {\emph {\bibinfo {title} {Topological Insulators:
				Dirac Equation in Condensed Matters}}}\ (\bibinfo  {publisher} {Springer},\
	\bibinfo {address} {Berlin, Germany},\ \bibinfo {year} {2012})\BibitemShut
	{NoStop}%
	\bibitem [{\citenamefont {Chiu}\ \emph {et~al.}(2016)\citenamefont {Chiu},
		\citenamefont {Teo}, \citenamefont {Schnyder},\ and\ \citenamefont
		{Ryu}}]{Chiu2016}%
	\BibitemOpen
	\bibfield  {author} {\bibinfo {author} {\bibfnamefont {C.-K.}\ \bibnamefont
			{Chiu}}, \bibinfo {author} {\bibfnamefont {J.~C.~Y.}\ \bibnamefont {Teo}},
		\bibinfo {author} {\bibfnamefont {A.~P.}\ \bibnamefont {Schnyder}}, \ and\
		\bibinfo {author} {\bibfnamefont {S.}~\bibnamefont {Ryu}},\ }\href {\doibase
		10.1103/RevModPhys.88.035005} {\bibfield  {journal} {\bibinfo  {journal}
			{Rev. Mod. Phys.}\ }\textbf {\bibinfo {volume} {88}},\ \bibinfo {pages}
		{035005} (\bibinfo {year} {2016})}\BibitemShut {NoStop}%
	\bibitem [{\citenamefont {Lindner}\ \emph {et~al.}(2011)\citenamefont
		{Lindner}, \citenamefont {Rafael},\ and\ \citenamefont {Galitski}}]{Lindner}%
	\BibitemOpen
	\bibfield  {author} {\bibinfo {author} {\bibfnamefont {N.~H.}\ \bibnamefont
			{Lindner}}, \bibinfo {author} {\bibfnamefont {G.}~\bibnamefont {Rafael}}, \
		and\ \bibinfo {author} {\bibfnamefont {V.}~\bibnamefont {Galitski}},\
	}\href@noop {} {\bibfield  {journal} {\bibinfo  {journal} {Nat. Phys.}\
		}\textbf {\bibinfo {volume} {7}},\ \bibinfo {pages} {490} (\bibinfo {year}
		{2011})}\BibitemShut {NoStop}%
	\bibitem [{\citenamefont {Cayssol}\ \emph {et~al.}(2013)\citenamefont
		{Cayssol}, \citenamefont {Dora}, \citenamefont {Simon},\ and\ \citenamefont
		{Moessner}}]{Cayssol13}%
	\BibitemOpen
	\bibfield  {author} {\bibinfo {author} {\bibfnamefont {J.}~\bibnamefont
			{Cayssol}}, \bibinfo {author} {\bibfnamefont {B.}~\bibnamefont {Dora}},
		\bibinfo {author} {\bibfnamefont {F.}~\bibnamefont {Simon}}, \ and\ \bibinfo
		{author} {\bibfnamefont {R.}~\bibnamefont {Moessner}},\ }\href@noop {}
	{\bibfield  {journal} {\bibinfo  {journal} {Phys. Status. Solidi. RRL}\
		}\textbf {\bibinfo {volume} {7}},\ \bibinfo {pages} {101} (\bibinfo {year}
		{2013})}\BibitemShut {NoStop}%
	\bibitem [{\citenamefont {Eckardt}(2017)}]{Eckardt17}%
	\BibitemOpen
	\bibfield  {author} {\bibinfo {author} {\bibfnamefont {A.}~\bibnamefont
			{Eckardt}},\ }\href@noop {} {\bibfield  {journal} {\bibinfo  {journal} {Rev.
				Mod. Phys.}\ }\textbf {\bibinfo {volume} {89}},\ \bibinfo {pages} {011004}
		(\bibinfo {year} {2017})}\BibitemShut {NoStop}%
	\bibitem [{\citenamefont {Oka}\ and\ \citenamefont {Kitamura}(2018)}]{Oka18}%
	\BibitemOpen
	\bibfield  {author} {\bibinfo {author} {\bibfnamefont {T.}~\bibnamefont
			{Oka}}\ and\ \bibinfo {author} {\bibfnamefont {S.}~\bibnamefont {Kitamura}},\
	}\href@noop {} {} (\bibinfo {year} {2018}),\ \bibinfo {note} {arXiv:
		1804.03212}\BibitemShut {NoStop}%
	\bibitem [{\citenamefont {Wang}\ \emph {et~al.}(2013)\citenamefont {Wang},
		\citenamefont {Steinberg}, \citenamefont {Jarillo-Herrero},\ and\
		\citenamefont {Gedik}}]{Wang13}%
	\BibitemOpen
	\bibfield  {author} {\bibinfo {author} {\bibfnamefont {Y.~H.}\ \bibnamefont
			{Wang}}, \bibinfo {author} {\bibfnamefont {H.}~\bibnamefont {Steinberg}},
		\bibinfo {author} {\bibfnamefont {P.}~\bibnamefont {Jarillo-Herrero}}, \ and\
		\bibinfo {author} {\bibfnamefont {N.}~\bibnamefont {Gedik}},\ }\href@noop {}
	{\bibfield  {journal} {\bibinfo  {journal} {Science}\ }\textbf {\bibinfo
			{volume} {342}},\ \bibinfo {pages} {453} (\bibinfo {year}
		{2013})}\BibitemShut {NoStop}%
	\bibitem [{\citenamefont {Maczewsky}\ \emph {et~al.}(2017)\citenamefont
		{Maczewsky}, \citenamefont {Zeuner}, \citenamefont {Nolte},\ and\
		\citenamefont {Szameit}}]{Maczewsky17}%
	\BibitemOpen
	\bibfield  {author} {\bibinfo {author} {\bibfnamefont {L.~J.}\ \bibnamefont
			{Maczewsky}}, \bibinfo {author} {\bibfnamefont {J.~M.}\ \bibnamefont
			{Zeuner}}, \bibinfo {author} {\bibfnamefont {S.}~\bibnamefont {Nolte}}, \
		and\ \bibinfo {author} {\bibfnamefont {A.}~\bibnamefont {Szameit}},\
	}\href@noop {} {\bibfield  {journal} {\bibinfo  {journal} {Nat. Commun.}\
		}\textbf {\bibinfo {volume} {8}},\ \bibinfo {pages} {13756} (\bibinfo {year}
		{2017})}\BibitemShut {NoStop}%
	\bibitem [{\citenamefont {Fleury}\ \emph {et~al.}(2016)\citenamefont {Fleury},
		\citenamefont {Khanikaev},\ and\ \citenamefont {Alu}}]{Fleury16}%
	\BibitemOpen
	\bibfield  {author} {\bibinfo {author} {\bibfnamefont {R.}~\bibnamefont
			{Fleury}}, \bibinfo {author} {\bibfnamefont {A.~B.}\ \bibnamefont
			{Khanikaev}}, \ and\ \bibinfo {author} {\bibfnamefont {A.}~\bibnamefont
			{Alu}},\ }\href@noop {} {\bibfield  {journal} {\bibinfo  {journal} {Nat.
				Commun.}\ }\textbf {\bibinfo {volume} {7}},\ \bibinfo {pages} {11744}
		(\bibinfo {year} {2016})}\BibitemShut {NoStop}%
	\bibitem [{\citenamefont {Ashcroft}\ and\ \citenamefont
		{Mermin}(1976)}]{Ashcroft_book}%
	\BibitemOpen
	\bibfield  {author} {\bibinfo {author} {\bibfnamefont {N.~W.}\ \bibnamefont
			{Ashcroft}}\ and\ \bibinfo {author} {\bibfnamefont {N.~D.}\ \bibnamefont
			{Mermin}},\ }\href@noop {} {\emph {\bibinfo {title} {Solid state physics}}}\
	(\bibinfo  {publisher} {Thomson Learning, Inc.},\ \bibinfo {address} {Boston,
		USA},\ \bibinfo {year} {1976})\BibitemShut {NoStop}%
	\bibitem [{\citenamefont {Chicone}(1999)}]{ODE_book}%
	\BibitemOpen
	\bibfield  {author} {\bibinfo {author} {\bibfnamefont {C.}~\bibnamefont
			{Chicone}},\ }\href@noop {} {\emph {\bibinfo {title} {Ordinary Differential
				Equations with Applications}}}\ (\bibinfo  {publisher} {Springer-Verlag},\
	\bibinfo {address} {New York},\ \bibinfo {year} {1999})\BibitemShut {NoStop}%
	\bibitem [{\citenamefont {Ladovrechis}\ and\ \citenamefont
		{Fulga}(2018)}]{Ladovrechis18}%
	\BibitemOpen
	\bibfield  {author} {\bibinfo {author} {\bibfnamefont {K.}~\bibnamefont
			{Ladovrechis}}\ and\ \bibinfo {author} {\bibfnamefont {I.~C.}\ \bibnamefont
			{Fulga}},\ }\href@noop {} {} (\bibinfo {year} {2018}),\ \bibinfo {note}
	{arXiv:1806.10099}\BibitemShut {NoStop}%
	\bibitem [{\citenamefont {Titum}\ \emph {et~al.}(2016)\citenamefont {Titum},
		\citenamefont {Berg}, \citenamefont {Rudner}, \citenamefont {Refael},\ and\
		\citenamefont {Lindner}}]{Titum}%
	\BibitemOpen
	\bibfield  {author} {\bibinfo {author} {\bibfnamefont {P.}~\bibnamefont
			{Titum}}, \bibinfo {author} {\bibfnamefont {E.}~\bibnamefont {Berg}},
		\bibinfo {author} {\bibfnamefont {M.~S.}\ \bibnamefont {Rudner}}, \bibinfo
		{author} {\bibfnamefont {G.}~\bibnamefont {Refael}}, \ and\ \bibinfo {author}
		{\bibfnamefont {N.~H.}\ \bibnamefont {Lindner}},\ }\href@noop {} {\bibfield
		{journal} {\bibinfo  {journal} {Phys. Rev. X}\ }\textbf {\bibinfo {volume}
			{6}},\ \bibinfo {pages} {021013} (\bibinfo {year} {2016})}\BibitemShut
	{NoStop}%
	\bibitem [{\citenamefont {Nathan}\ \emph {et~al.}(2017)\citenamefont {Nathan},
		\citenamefont {Rudner}, \citenamefont {Lindner}, \citenamefont {Berg},\ and\
		\citenamefont {Refael}}]{Nathan}%
	\BibitemOpen
	\bibfield  {author} {\bibinfo {author} {\bibfnamefont {F.}~\bibnamefont
			{Nathan}}, \bibinfo {author} {\bibfnamefont {M.~S.}\ \bibnamefont {Rudner}},
		\bibinfo {author} {\bibfnamefont {N.~H.}\ \bibnamefont {Lindner}}, \bibinfo
		{author} {\bibfnamefont {E.}~\bibnamefont {Berg}}, \ and\ \bibinfo {author}
		{\bibfnamefont {G.}~\bibnamefont {Refael}},\ }\href@noop {} {\bibfield
		{journal} {\bibinfo  {journal} {Phys. Rev. Lett}\ }\textbf {\bibinfo {volume}
			{119}},\ \bibinfo {pages} {186801} (\bibinfo {year} {2017})}\BibitemShut
	{NoStop}%
	\bibitem [{\citenamefont {Fulga}\ and\ \citenamefont
		{Maksymenko}(2016)}]{Fulga16}%
	\BibitemOpen
	\bibfield  {author} {\bibinfo {author} {\bibfnamefont {I.~C.}\ \bibnamefont
			{Fulga}}\ and\ \bibinfo {author} {\bibfnamefont {M.}~\bibnamefont
			{Maksymenko}},\ }\href@noop {} {\bibfield  {journal} {\bibinfo  {journal}
			{Phys. Rev. B}\ }\textbf {\bibinfo {volume} {93}},\ \bibinfo {pages} {075405}
		(\bibinfo {year} {2016})}\BibitemShut {NoStop}%
	\bibitem [{\citenamefont {Tauber}(2018)}]{Tauber18}%
	\BibitemOpen
	\bibfield  {author} {\bibinfo {author} {\bibfnamefont {C.}~\bibnamefont
			{Tauber}},\ }\href@noop {} {\bibfield  {journal} {\bibinfo  {journal} {Phys.
				Rev. B}\ }\textbf {\bibinfo {volume} {97}},\ \bibinfo {pages} {195312}
		(\bibinfo {year} {2018})}\BibitemShut {NoStop}%
	\bibitem [{\citenamefont {Hainaut}\ \emph {et~al.}(2018)\citenamefont
		{Hainaut}, \citenamefont {Rancon}, \citenamefont {Clement}, \citenamefont
		{Manai}, \citenamefont {Szriftgiser}, \citenamefont {Delande}, \citenamefont
		{Garreau},\ and\ \citenamefont {Chicireanu}}]{Hainaut18}%
	\BibitemOpen
	\bibfield  {author} {\bibinfo {author} {\bibfnamefont {C.}~\bibnamefont
			{Hainaut}}, \bibinfo {author} {\bibfnamefont {A.}~\bibnamefont {Rancon}},
		\bibinfo {author} {\bibfnamefont {J.~F.}\ \bibnamefont {Clement}}, \bibinfo
		{author} {\bibfnamefont {I.}~\bibnamefont {Manai}}, \bibinfo {author}
		{\bibfnamefont {P.}~\bibnamefont {Szriftgiser}}, \bibinfo {author}
		{\bibfnamefont {D.}~\bibnamefont {Delande}}, \bibinfo {author} {\bibfnamefont
			{J.~C.}\ \bibnamefont {Garreau}}, \ and\ \bibinfo {author} {\bibfnamefont
			{R.}~\bibnamefont {Chicireanu}},\ }\href@noop {} {} (\bibinfo {year}
	{2018}),\ \bibinfo {note} {arXiv: 1811.10244}\BibitemShut {NoStop}%
	\bibitem [{\citenamefont {Prodan}\ and\ \citenamefont
		{Schulz-Baldes}(2016)}]{Prodan_book}%
	\BibitemOpen
	\bibfield  {author} {\bibinfo {author} {\bibfnamefont {E.}~\bibnamefont
			{Prodan}}\ and\ \bibinfo {author} {\bibfnamefont {H.}~\bibnamefont
			{Schulz-Baldes}},\ }\href@noop {} {\emph {\bibinfo {title} {Bulk and boundary
				invariants for complex topological insulators: From K-theory to physics}}}\
	(\bibinfo  {publisher} {Springer International Publishing},\ \bibinfo
	{address} {Switzerland},\ \bibinfo {year} {2016})\BibitemShut {NoStop}%
	\bibitem [{\citenamefont {Rudner}\ \emph {et~al.}(2013)\citenamefont {Rudner},
		\citenamefont {Lindner}, \citenamefont {Berg},\ and\ \citenamefont
		{Levin}}]{Rudner}%
	\BibitemOpen
	\bibfield  {author} {\bibinfo {author} {\bibfnamefont {M.~S.}\ \bibnamefont
			{Rudner}}, \bibinfo {author} {\bibfnamefont {N.~H.}\ \bibnamefont {Lindner}},
		\bibinfo {author} {\bibfnamefont {E.}~\bibnamefont {Berg}}, \ and\ \bibinfo
		{author} {\bibfnamefont {M.}~\bibnamefont {Levin}},\ }\href@noop {}
	{\bibfield  {journal} {\bibinfo  {journal} {Phys. Rev. X}\ }\textbf {\bibinfo
			{volume} {3}},\ \bibinfo {pages} {031005} (\bibinfo {year}
		{2013})}\BibitemShut {NoStop}%
	\bibitem [{\citenamefont {Sadel}\ and\ \citenamefont
		{Schultz-Baldes}(2017)}]{Sadel17}%
	\BibitemOpen
	\bibfield  {author} {\bibinfo {author} {\bibfnamefont {C.}~\bibnamefont
			{Sadel}}\ and\ \bibinfo {author} {\bibfnamefont {H.}~\bibnamefont
			{Schultz-Baldes}},\ }\href@noop {} {\bibfield  {journal} {\bibinfo  {journal}
			{Math. Phys. Anal. Geom.}\ }\textbf {\bibinfo {volume} {20}},\ \bibinfo
		{pages} {22} (\bibinfo {year} {2017})}\BibitemShut {NoStop}%
	\bibitem [{\citenamefont {Graf}\ and\ \citenamefont {Tauber}(2018)}]{Graf18}%
	\BibitemOpen
	\bibfield  {author} {\bibinfo {author} {\bibfnamefont {G.~M.}\ \bibnamefont
			{Graf}}\ and\ \bibinfo {author} {\bibfnamefont {C.}~\bibnamefont {Tauber}},\
	}\href@noop {} {\bibfield  {journal} {\bibinfo  {journal} {Ann. Henri
				Poincare}\ }\textbf {\bibinfo {volume} {19}},\ \bibinfo {pages} {709}
		(\bibinfo {year} {2018})}\BibitemShut {NoStop}%
	\bibitem [{Ban()}]{Bandnote}%
	\BibitemOpen
	\href@noop {} {}\bibinfo {note} {Note that for the FTI, the definitions of
		the top and bottom bands are completely artificial because it depends on how
		one cuts the circle of the quasi-energy spectrum into a finite
		segment.}\BibitemShut {Stop}%
	\bibitem [{\citenamefont {Kennedy}\ and\ \citenamefont
		{Guggenheim}(2015)}]{Kennedy15}%
	\BibitemOpen
	\bibfield  {author} {\bibinfo {author} {\bibfnamefont {R.}~\bibnamefont
			{Kennedy}}\ and\ \bibinfo {author} {\bibfnamefont {C.}~\bibnamefont
			{Guggenheim}},\ }\href@noop {} {\bibfield  {journal} {\bibinfo  {journal}
			{Phys. Rev. B}\ }\textbf {\bibinfo {volume} {91}},\ \bibinfo {pages} {245148}
		(\bibinfo {year} {2015})}\BibitemShut {NoStop}%
	\bibitem [{\citenamefont {Puttmann}\ and\ \citenamefont
		{Rigas}(2003)}]{Puttmann03}%
	\BibitemOpen
	\bibfield  {author} {\bibinfo {author} {\bibfnamefont {T.}~\bibnamefont
			{Puttmann}}\ and\ \bibinfo {author} {\bibfnamefont {A.}~\bibnamefont
			{Rigas}},\ }\href@noop {} {\bibfield  {journal} {\bibinfo  {journal}
			{Comment. Math. Helv.}\ }\textbf {\bibinfo {volume} {78}},\ \bibinfo {pages}
		{648} (\bibinfo {year} {2003})}\BibitemShut {NoStop}%
	\bibitem [{\citenamefont {Carpentier}\ \emph {et~al.}(2015)\citenamefont
		{Carpentier}, \citenamefont {Delplace}, \citenamefont {Fruchart},\ and\
		\citenamefont {Gawedzki}}]{Carpentier15}%
	\BibitemOpen
	\bibfield  {author} {\bibinfo {author} {\bibfnamefont {D.}~\bibnamefont
			{Carpentier}}, \bibinfo {author} {\bibfnamefont {P.}~\bibnamefont
			{Delplace}}, \bibinfo {author} {\bibfnamefont {M.}~\bibnamefont {Fruchart}},
		\ and\ \bibinfo {author} {\bibfnamefont {K.}~\bibnamefont {Gawedzki}},\
	}\href@noop {} {\bibfield  {journal} {\bibinfo  {journal} {Phys. Rev. Lett.}\
		}\textbf {\bibinfo {volume} {114}},\ \bibinfo {pages} {106806} (\bibinfo
		{year} {2015})}\BibitemShut {NoStop}%
	\bibitem [{\citenamefont {Nathan}\ and\ \citenamefont
		{Rudner}(2015)}]{Nathan15}%
	\BibitemOpen
	\bibfield  {author} {\bibinfo {author} {\bibfnamefont {F.}~\bibnamefont
			{Nathan}}\ and\ \bibinfo {author} {\bibfnamefont {M.~S.}\ \bibnamefont
			{Rudner}},\ }\href@noop {} {\bibfield  {journal} {\bibinfo  {journal} {New J.
				Phys.}\ }\textbf {\bibinfo {volume} {17}},\ \bibinfo {pages} {125014}
		(\bibinfo {year} {2015})}\BibitemShut {NoStop}%
	\bibitem [{\citenamefont {Fruchart}(2016)}]{Fruchart16}%
	\BibitemOpen
	\bibfield  {author} {\bibinfo {author} {\bibfnamefont {M.}~\bibnamefont
			{Fruchart}},\ }\href@noop {} {\bibfield  {journal} {\bibinfo  {journal}
			{Phys. Rev. B}\ }\textbf {\bibinfo {volume} {93}},\ \bibinfo {pages} {115429}
		(\bibinfo {year} {2016})}\BibitemShut {NoStop}%
	\bibitem [{\citenamefont {Roy}\ and\ \citenamefont {Harper}(2017)}]{Roy}%
	\BibitemOpen
	\bibfield  {author} {\bibinfo {author} {\bibfnamefont {R.}~\bibnamefont
			{Roy}}\ and\ \bibinfo {author} {\bibfnamefont {F.}~\bibnamefont {Harper}},\
	}\href@noop {} {\bibfield  {journal} {\bibinfo  {journal} {Phys. Rev. B}\
		}\textbf {\bibinfo {volume} {96}},\ \bibinfo {pages} {155118} (\bibinfo
		{year} {2017})}\BibitemShut {NoStop}%
	\bibitem [{\citenamefont {Qi}\ \emph {et~al.}(2008)\citenamefont {Qi},
		\citenamefont {Hughes},\ and\ \citenamefont {Zhang}}]{Qi2}%
	\BibitemOpen
	\bibfield  {author} {\bibinfo {author} {\bibfnamefont {X.~L.}\ \bibnamefont
			{Qi}}, \bibinfo {author} {\bibfnamefont {T.~L.}\ \bibnamefont {Hughes}}, \
		and\ \bibinfo {author} {\bibfnamefont {S.~C.}\ \bibnamefont {Zhang}},\
	}\href@noop {} {\bibfield  {journal} {\bibinfo  {journal} {Phys. Rev. B}\
		}\textbf {\bibinfo {volume} {78}},\ \bibinfo {pages} {195424} (\bibinfo
		{year} {2008})}\BibitemShut {NoStop}%
	\bibitem [{\citenamefont {Bott}\ and\ \citenamefont {Tu}(1982)}]{Bott_book}%
	\BibitemOpen
	\bibfield  {author} {\bibinfo {author} {\bibfnamefont {R.}~\bibnamefont
			{Bott}}\ and\ \bibinfo {author} {\bibfnamefont {L.~W.}\ \bibnamefont {Tu}},\
	}\href@noop {} {\emph {\bibinfo {title} {Differential Forms in Algebraic
				Topology}}}\ (\bibinfo  {publisher} {Springer-Verlag},\ \bibinfo {address}
	{New York},\ \bibinfo {year} {1982})\BibitemShut {NoStop}%
	\bibitem [{\citenamefont {Witten}(1983{\natexlab{a}})}]{Witten}%
	\BibitemOpen
	\bibfield  {author} {\bibinfo {author} {\bibfnamefont {E.}~\bibnamefont
			{Witten}},\ }\href@noop {} {\bibfield  {journal} {\bibinfo  {journal} {Nucl.
				Phys. B}\ }\textbf {\bibinfo {volume} {223}},\ \bibinfo {pages} {422}
		(\bibinfo {year} {1983}{\natexlab{a}})}\BibitemShut {NoStop}%
	\bibitem [{\citenamefont {Witten}(1983{\natexlab{b}})}]{Witten1}%
	\BibitemOpen
	\bibfield  {author} {\bibinfo {author} {\bibfnamefont {E.}~\bibnamefont
			{Witten}},\ }\href@noop {} {\bibfield  {journal} {\bibinfo  {journal} {Nucl.
				Phys. B}\ }\textbf {\bibinfo {volume} {223}},\ \bibinfo {pages} {433}
		(\bibinfo {year} {1983}{\natexlab{b}})}\BibitemShut {NoStop}%
	\bibitem [{\citenamefont {Gawedzki}(2017)}]{Gawedzki17}%
	\BibitemOpen
	\bibfield  {author} {\bibinfo {author} {\bibfnamefont {K.}~\bibnamefont
			{Gawedzki}},\ }\href@noop {} {\bibfield  {journal} {\bibinfo  {journal} {J.
				Geom. Phys.}\ }\textbf {\bibinfo {volume} {120}},\ \bibinfo {pages} {169}
		(\bibinfo {year} {2017})}\BibitemShut {NoStop}%
	\bibitem [{\citenamefont {Monaco}\ and\ \citenamefont
		{Tauber}(2017)}]{Monaco17}%
	\BibitemOpen
	\bibfield  {author} {\bibinfo {author} {\bibfnamefont {D.}~\bibnamefont
			{Monaco}}\ and\ \bibinfo {author} {\bibfnamefont {C.}~\bibnamefont
			{Tauber}},\ }\href@noop {} {\bibfield  {journal} {\bibinfo  {journal} {Lett.
				Math. Phys.}\ }\textbf {\bibinfo {volume} {107}},\ \bibinfo {pages} {1315}
		(\bibinfo {year} {2017})}\BibitemShut {NoStop}%
	\bibitem [{\citenamefont {Fukui}\ \emph {et~al.}(2008)\citenamefont {Fukui},
		\citenamefont {Fujiwara},\ and\ \citenamefont {Hatsugai}}]{Hatsugai}%
	\BibitemOpen
	\bibfield  {author} {\bibinfo {author} {\bibfnamefont {T.}~\bibnamefont
			{Fukui}}, \bibinfo {author} {\bibfnamefont {T.}~\bibnamefont {Fujiwara}}, \
		and\ \bibinfo {author} {\bibfnamefont {Y.}~\bibnamefont {Hatsugai}},\
	}\href@noop {} {\bibfield  {journal} {\bibinfo  {journal} {J. Phys. Soc.
				Jpn.}\ }\textbf {\bibinfo {volume} {77}},\ \bibinfo {pages} {123705}
		(\bibinfo {year} {2008})}\BibitemShut {NoStop}%
	\bibitem [{\citenamefont {Nakahara}(2003)}]{Nakahara_book}%
	\BibitemOpen
	\bibfield  {author} {\bibinfo {author} {\bibfnamefont {M.}~\bibnamefont
			{Nakahara}},\ }\href@noop {} {\emph {\bibinfo {title} {Geometry, Topology and
				Physics, second edition}}}\ (\bibinfo  {publisher} {Institute of Physics
		Publishing},\ \bibinfo {address} {Bristol},\ \bibinfo {year}
	{2003})\BibitemShut {NoStop}%
	\bibitem [{\citenamefont {Nair}(2005)}]{Nair}%
	\BibitemOpen
	\bibfield  {author} {\bibinfo {author} {\bibfnamefont {V.~P.}\ \bibnamefont
			{Nair}},\ }\href@noop {} {\emph {\bibinfo {title} {Quantum Field Theory: A
				Modern Perspective}}}\ (\bibinfo  {publisher} {Springer},\ \bibinfo {address}
	{New York, USA},\ \bibinfo {year} {2005})\BibitemShut {NoStop}%
	\bibitem [{\citenamefont {Hu}(1959)}]{S-T-Hu}%
	\BibitemOpen
	\bibfield  {author} {\bibinfo {author} {\bibfnamefont {S.-T.}\ \bibnamefont
			{Hu}},\ }\href@noop {} {\emph {\bibinfo {title} {Homotopy Theory}}}\
	(\bibinfo  {publisher} {Acdemic Press, Inc},\ \bibinfo {address} {New York
		and London},\ \bibinfo {year} {1959})\BibitemShut {NoStop}%
	\bibitem [{\citenamefont {Deng}\ \emph {et~al.}(2013)\citenamefont {Deng},
		\citenamefont {Wang}, \citenamefont {Shen},\ and\ \citenamefont
		{Duan}}]{Duan}%
	\BibitemOpen
	\bibfield  {author} {\bibinfo {author} {\bibfnamefont {D.~L.}\ \bibnamefont
			{Deng}}, \bibinfo {author} {\bibfnamefont {S.~T.}\ \bibnamefont {Wang}},
		\bibinfo {author} {\bibfnamefont {C.}~\bibnamefont {Shen}}, \ and\ \bibinfo
		{author} {\bibfnamefont {L.~M.}\ \bibnamefont {Duan}},\ }\href@noop {}
	{\bibfield  {journal} {\bibinfo  {journal} {Phys. Rev. B}\ }\textbf {\bibinfo
			{volume} {88}},\ \bibinfo {pages} {201105(R)} (\bibinfo {year}
		{2013})}\BibitemShut {NoStop}%
	\bibitem [{\citenamefont {Nur~Unal}\ \emph {et~al.}(2018)\citenamefont
		{Nur~Unal}, \citenamefont {Seradjeh},\ and\ \citenamefont
		{Eckardt}}]{Unal18}%
	\BibitemOpen
	\bibfield  {author} {\bibinfo {author} {\bibfnamefont {F.}~\bibnamefont
			{Nur~Unal}}, \bibinfo {author} {\bibfnamefont {B.}~\bibnamefont {Seradjeh}},
		\ and\ \bibinfo {author} {\bibfnamefont {A.}~\bibnamefont {Eckardt}},\
	}\href@noop {} {} (\bibinfo {year} {2018}),\ \bibinfo {note}
	{arXiv:1812.04636}\BibitemShut {NoStop}%
\end{thebibliography}
%

\end{document}